\documentclass[a4paper,11pt]{article}
\pdfoutput=1 

\usepackage{jinstpub} 

\usepackage{enumitem} 

\title{\boldmath A fast numerical method for photomultiplier tube calibration} 
\author[a,1]{L. N. Kalousis,\note{Corresponding author.}}
\author[a]{J. P. A. M. de Andr\'e,}
\author[a]{E. Baussan}
\author[a]{and M. Dracos}

\affiliation[a]{IPHC, Universit\'e de Strasbourg, CNRS/IN2P3, F-67037 Strasbourg, France}
\emailAdd{leonidas.kalousis@iphc.cnrs.fr}

\abstract{In this article, a new method is discussed for the calibration and monitoring of photomultiplier tubes (PMTs). 
This method is based on a Discrete Fourier Transform (DFT) and it is fast and general so that it can be used in cases where an analytical model of the PMT response is not available. 
The DFT approach is employed for the absolute calibration of the Hamamatsu R1408 photomultiplier tube. 
It should be noted that the R1408 PMTs do not show a sharp peak at the single photoelectron distribution and gain determination via conventional methods is often unattainable. 
Here, we show that the DFT technique, coupled with a gamma function model for the single photoelectron response, produces rigorous calibration results  and it can be used for gain determination with a good accuracy.}


\arxivnumber{1911.06220}

\begin{document}
\maketitle
\flushbottom

\section{Introduction and outline}
\label{sec:intro}

Photomultiplier tubes (PMTs) are devices extensively used in low-energy nuclear physics, particle physics and medical applications~\cite{Leo}. 
Their striking stability, simple principles of operation and relatively low cost make them attractive solutions in the instrumentation of large coverage detectors. 
Within neutrino physics, massive monolithic detectors usually employ a sizable number of PMTs to detect optical signals created in the target material. 
For example, the Super-Kamiokande detector utilizes around 11,000 20'' PMTs, that are mounted on the walls of a cylindrical tank containing 50~kt of ultra pure water~\cite{SK}. 
Another example is the medium-baseline reactor experiment JUNO that is expected to use $\sim$18,000 20'' PMTs in the Central Detector,
in order to achieve the unprecedented energy resolution of $3\%/ \sqrt{E_\nu}$ that is necessary for an unambiguous determination of the neutrino mass ordering. 
Additionally, JUNO will further use $\sim$25,000 3'' PMTs in the Central Detector to improve the energy scale calibration and 
$\sim$2,000 20'' PMTs in the surrounding water Cherenkov Outer Veto for the tagging of cosmic muons~\cite{JUNO1,JUNO2,JUNO3}. 

A very important aspect of PMT's operation is the fact that the number of photoelectrons (PEs) collected and focused in the first stage of amplification is 
(up to expected statistical fluctuations) proportional to the measured charge at the output of the dynode chain.\footnote{%
The basic operating principles of PMTs are reviewed in ref.~\cite{Leo} and will not be repeated here.}
The proportionality factor that relates charge and PEs is formally called \emph{gain} and it is perhaps the most significant parameter of a PMT. 
Furthermore, it should be emphasized that the number of PEs detected, usually, depends on the energy released by a particle inside the detector. 
Thus, the gain value is necessary for deducing the energy of incident particles in a detector instrumented with PMTs.   

In section~\ref{sec:deconv}, we lay down in some details the standard technique for gain determination of PMTs since it is so critical in the understanding of the material included in this publication. 
In doing so, we introduce the notation employed throughout the whole document. 
In section~\ref{sec:dft} we propose a novel, high precision numerical approach that can be used to calculate the PMT charge response function, $S_R(x)$. 
The proposed method leaning on the Discrete Fourier Transform (DFT) is general enough that can be exploited whenever a closed form of $S_R(x)$ (or even a valid approximation) can not be derived.
This implementation is much faster than a direct numerical calculation of convolutions owing to the efficient DFT numerical algorithms that are available. 
Finally, in section~\ref{sec:results} we demonstrate the utility of this new approach through the absolute calibration of the Hamamatsu R1408 model PMT in a rigorous and precise manner.

\section{Gain determination method}
\label{sec:deconv}

The standard analysis method used for PMT gain determination was first presented in an influential paper written by E. H. Bellamy and collaborators in 1994~\cite{Bellamy}. 
In this work, a refined procedure is given that can deconvolute the charge distribution of a PMT in the single photoelectron mode through a simple but sophisticated statistical analysis of PMT spectra from a pulsed light source. 
This operation gives access to the main parameters of the process, such as the PMT gain and the mean number of PEs recorded.

What makes this approach appealing is that the actual knowledge of the light source characteristics is not at all essential to the method, these characteristics (e.g. light intensity) should only be stable during time.  
The only ingredient required is a mathematical scheme that models, in a realistic way, the response spectra of the PMTs when illuminated with such light pulses. 
The model can then be applied to real spectra taken under these conditions and extract both the gain and all the relevant attributes of the light source with the use of a minimization package. 
In the remaining parts of this section, we describe the main ingredients of the Bellamy et al.~\cite{Bellamy} gain determination method since it constitutes the embarkation point of the DFT approach presented in section~\ref{sec:dft}. 

\subsection{Single photoelectron charge amplification}

In modeling the response of a PMT, one is usually forced to postulate a distribution function for the single photoelectron (SPE) charge deposition. 
The calculation from first principles of this distribution, taking into account all multiplication stages, is very complicated and assumes a very good knowledge of all multiplication parameters. 
That is, when a PE enters the multiplicative dynode structure, the law for the charge amplification is not deterministic but, on the contrary, follows from a probability distribution function (PDF). 
This distribution is characteristic of the PMT and the voltage shared among its dynodes. In what follows it shall be termed $S(x)$. 

$S(x)dx$ gives the probability to collect an amount of charge between $x$ and $x + dx$ in the output of the PMT whenever a single PE
is released from the photocathode and focused in the dynode system. 
$S(x)$ is normalized in the usual sense:
\begin{align}
\int_{0}^{+\infty} S(x) \ dx = 1. \label{eq:norm}
\end{align}
It is further assumed that its statistical moments are all well defined. 
Note that the integration in eq.~\eqref{eq:norm} starts from zero since the charge is taken, by definition, to be positive. 
In particular, in this formalism, a PMT cannot produce negative charges in response to a light source. 

In the process of PMT calibration, different types of PDFs are used depending, each time, on the PMT under examination.
For example, the R7081 Hamamatsu PMTs can be adequately described by a gaussian plus an exponential part~\cite{DCID1,DCID2,DCID3}. 
On the other hand, the R1408 Hamamatsu PMT model is effectively parameterized through a gamma distribution~\cite{Me}. 
We should note that for all the studies conducted for the purposes of this article we were always under the PMT's saturation point. 
In this case, the charge deposition is a linear process with respect to the light intensity. 

\subsection{Poisson photoelectron production}

Whenever light pulses, created for example by a laser or a light emitting diode, hit the photocathode, 
there is a certain probability for electrons to be produced (external photoelectric effect).
This process is related to the quantum efficiency 
of the photocathode~\cite{Hama}. 
Since the number of photons in each pulse is roughly constant and the probability for photoconversion follows from a random distribution, 
the final probability for $n$ PEs to be created is governed by a Poisson distribution~\cite{Zorin}: 
\begin{align}
P( n; \mu ) =  \frac{\mu^n}{n!} e^{-\mu}, \label{eq:poisson}
\end{align}
with $\mu= <n_\gamma> \epsilon_q$. 
$<n_\gamma>$ and $\epsilon_q$ are the mean number of photons of the light pulses and the quantum efficiency of the PMT respectively. 
It is worth noting that the probability the light pulse will produce no electrons is different from zero and in fact equals to $P(0;\mu) = e^{-\mu}$.
That is, just by counting the ``zero'' cases gives a direct access to the mean value $\mu$. 

\begin{figure}[!t]
\centering
\includegraphics[width=9.0cm, height=7.0cm]{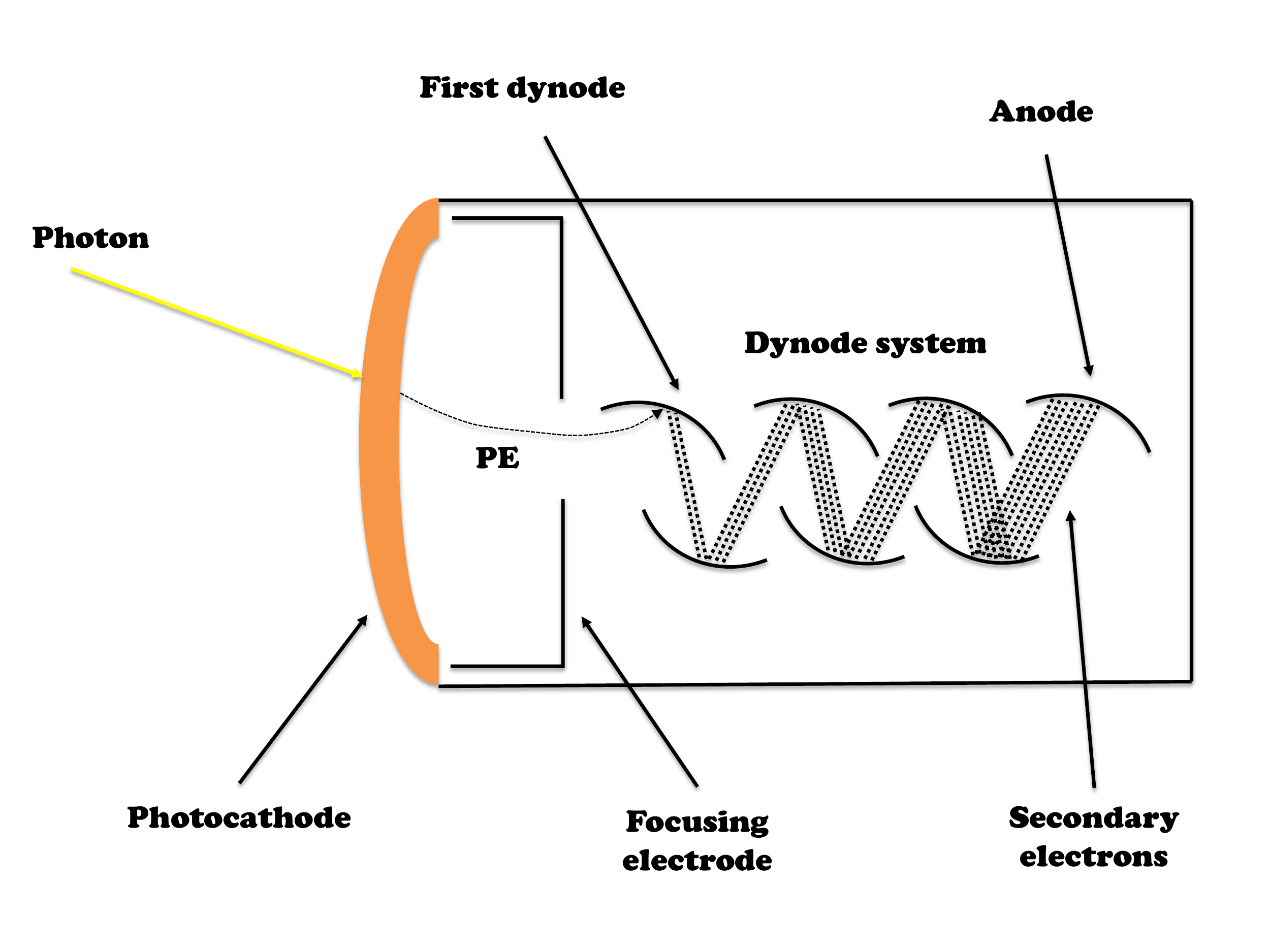} 
\caption{Schematic of the PMT operation principle.
A photon converts on the photocathode to produce a single photoelectron that is directed on the dynode chain for multiplication. }
\label{fig:pm}
\end{figure}

After photoconversion, the electrons are accelerated, focused, and finally subjected to the multiplicative dynode structure. 
It is a well-known fact that not all of the released PEs survive this secondary process and finally enter the amplification chain. 
The probability for a PE to reach the first dynode is called collection efficiency and usually denoted as $\eta$. 
This secondary collection process follows a random binary distribution that modifies the initial Poisson mean by $\mu \rightarrow \mu^\prime = \mu \cdot \eta$. 
In what follows we shall use the symbol $\mu$ in the place of $\mu^\prime$.  
Nevertheless, we must never forget, that this $\mu$ is the modified mean according to the collection efficiency and thus characterizes jointly both the light pulse and the quantum and collection efficiencies.
Figure~\ref{fig:pm} shows the basic steps of a PMT operation (photoconversion, focusing, multiplication, etc.) in a simple schematic. 

\enlargethispage{-\baselineskip}
Let us now consider the following question: 
what will be the combined probability for a light pulse, described by eq.~\eqref{eq:poisson}, to produce a single PE that will afterwards deposit an amount of charge between $x$ and $x+dx$ ? 
The answer is straightforward from our previous analysis. The total probability is just the product of the two separate probabilities.
\begin{align}
P(1;\mu) S(x) dx 
\end{align}
Let us now repeat this question but for the case of a two PE emission. 
So, what will be the combined probability for this light pulse to produce two PEs that will afterwards deposit an amount of charge between $x$ and $x + dx$ ? 
The answer turns out to be not that obvious.

The total probability for the first PE to deposit a charge between $y$ and $y$ + $dy$, 
and the second PE charge between $z$ and $z$ + $dz$ is given by the product of the two separate  probabilities since the processes are fully independent: 
\begin{align}
S(y)dy \ S(z)dz.
\end{align}
Denoting as $x = y+z$ the total charge recorded in the PMT anode, this probability can by rewritten in the form: 
\begin{align}
S(x-z) dx \ S(z)dz
\end{align}
by means of simple substitution. 
To obtain the probability for a total charge deposition between $x$ and $x$ + $dx$ one has to integrate over all $z$ charges. 
\begin{align}
\left( \ \int_{0}^{+\infty} S(x-z) \ S(z)dz \ \right) dx 
\end{align}
The expression in the parenthesis is the convolution of $S(x)$ with itself. 
Thus, the two PEs produce a total charge $x$, according to the probability distribution $S_2(x)$ where:
\begin{align}
S_2(x) = (S * S )(x).
\end{align}
The answer to our second question can be written down and is a simple generalization of our first result.
\begin{align}
P(2;\mu) S_2(x) dx 
\end{align}
Of course, similar arguments hold for the case of three PEs, four PEs, and so forth. 

We recognize that the total probability for a light pulse to produce charge between $x$ and $x+dx$ is given by: 
\begin{align}
\left( \ \sum_{n=0}^{+\infty} P( n; \mu ) S_n(x) \ \right) dx,
\label{eq:Sid}
\end{align}
where eq.~\eqref{eq:Sid} is the sum of the individual probabilities for any given number of PEs.
$S_n(x)$ is the $n$-times convolution of $S(x)$ and $S_0(x)$ the zero PE charge distribution. 
$S_0(x)$ is the delta function $\delta(x)$, that is if zero PEs are produced all charge is deposited at zero.  
In general, $S_n(x)$ is given by:
\begin{align}
S_n(x)  = \begin{cases}
   \ \quad \delta (x),       & \text{for}  \ n = 0 \\
    (S*S_{n-1})(x),  & \text{for} \ n \geqslant 1 .
  \end{cases}
\end{align}
We note that for $n=1$,  $S_1(x)$ is equal to $S(x)$ since $ (S * \delta)(x) = S(x) $.

\subsection{Incorporating the pedestal}

The PDF for the PMT charge output in response to a poissonian light source is named $S_{ID}(x)$. 
The form of $S_{ID}(x)$ was worked out in the previous subsection and it is repeated here:
\begin{align}
S_{ID}(x) = \sum_{n=0}^{+\infty} P( n; \mu ) S_n(x). 
\label{eq:Sid2}
\end{align}
We choose to call this PDF $S_{ID}(x)$ because it is ideal in the sense that it cannot account for any external charge fluctuations as those produced by electronic noise.
Indeed, never in our previous analysis did we mention about the PMT pedestal and the charge smearing that's responsible for. 

To incorporate the pedestal background charge in this formalism, a shifted gaussian over a mean value $Q_0$ is used. 
\begin{align}
B(x) = \frac{1}{\sqrt{2\pi}\sigma_0} e^{ - \frac{( x - Q_0 )^2}{2\sigma_0^2}}
\end{align}
$\sigma_0$ describes the charge smearing induced by the electronic noise.
For the same reasons we needed to convolute the signal distribution $S(x)$ with itself to obtain the two PE $S_2(x)$ distribution, 
here we need to convolute the $S_{ID}(x)$ and $B(x)$ distributions to produce the output signal PDF $S_{R}(x)$. 
\begin{align}
S_R(x) = &     (S_{ID}*B)(x) \nonumber  \\
            = & \sum_{n=0}^{+\infty} P( n; \mu ) ( S_n * B )(x). 
\label{eq:Sr0}
\end{align}
Here ``R'' stands for real. 
Note that the position of the baseline of the signal depends on the setup and it's not always set at zero.  This implies that the mean charge of the pedestal
($Q_0$, the integral of the baseline when no PEs have been detected), is not exactly at zero unless a baseline subtraction has been attempted. 
The introduction of the pedestal through this last convolution has the effect of shifting $S_{ID}(x)$ by $Q_0$ and smearing the various PE peaks. 
In general, our job is to derive $S_R(x)$ in a closed or, at least, in an approximate but still useful form.

\subsection{Low $\mu$ approximation}

At first this might seem like an extraordinary enterprise but a valid approximation can be easily derived if the mean value of the Poisson distribution ($\mu$) is low enough. 
In that case, if $\mu$ is small $S_R(x)$ can be written as: 
\begin{align}
S_{R}(x) \simeq \sum_{n=0}^{m} P( n; \mu ) (S_n*B)(x), 
\label{eq:Sr}
\end{align}
where $m$ is an integer chosen such that the $ P( n; \mu ) (S_n*B)(x)$ terms with $n > m $ are negligible owing to the dumping of the Poisson factors for $n > m $. 
When $\mu= 1$ then $m$ can be chosen to be equal to five, while when $\mu= 2$ the first nine terms of eq.~\eqref{eq:Sr} are enough to give a precision of better than 1\% in both cases.

The aforementioned recipe was first presented in ref.~\cite{Bellamy}.  
In that paper, the authors give an exact formula for the $(S_n*B)(x)$ distributions when the SPE response function $S(x)$ is given by a gaussian. 
$S_R(x)$ can then be used to fit real data taken while the PMT is illuminated with low-intensity light pulses and the parameters of $P(n;\mu)$, $B(x)$ and $S(x)$ can be extracted. 
The PMT gain $G$ is then given by: 
\begin{align}
G =  \int_{0}^{+\infty} x S(x) \ dx,
\end{align}
which is the mean value of $S(x)$.

On the other hand, the PMT model treated in ref.~\cite{Smirnov} is better described by a combination of a truncated gaussian plus an exponential distribution:
\begin{align}
S(x) =    \left( \ w \alpha e^{-\alpha x } + \frac{(1-w)}{g_N} \frac{1}{\sqrt{2\pi}\sigma_1} e^{ - \frac{( x - Q_1 )^2}{2\sigma_1^2}} \ \right) H( x )
\label{eq:smi}
\end{align}
with 
\begin{align}
g_N =  \frac{1}{2} \left(  \    1 +  \text{erf} \left( \frac{Q_1}{\sqrt{2}\sigma_1 } \right) \   \right). 
\end{align}
The Heaviside $H(x)$ step function is necessary because negative charges are not physical and $g_N$ is defined so that $S(x)$ is normalized to one. 
In this case, $S_R(x)$ does not have a simple analytic expression but an approximate formula is given in ref.~\cite{Smirnov}. 
An example of a single fit using this gaussian model can be found in ref.~\cite{darkside}, figure~2. Note that in this case the gain is given by the mean of Eq.~\eqref{eq:smi}. 
Furthermore, in ref.~\cite{darkside} 
the one and two PE contributions to $S_R(x)$ are calculated numerically. 

For more complicated cases, as for example the position sensitive Hamamatsu H8500C-03 PMTs, a rather sophisticated model has been presented in ref.~\cite{pavel}.
However, in general, it becomes burdensome to calculate the contributions from additional PEs due to the brute force calculation of the convolutions in the $S_R(x)$ formula. 
It is the purpose of this paper is to offer a novel and efficient numerical method that can be used to evaluate $S_R(x)$ 
whenever an analytical form of $(S_n*B)(x)$ functions is not  achievable. 
This method works for any generic $S(x)$ distribution and is capable of providing $S_R(x)$ for any value of the poissonian mean $\mu$.

\section{Discrete Fourier Transform approach to $S_R(x)$}
\label{sec:dft}

To start an exposition of the numerical method developed for the purposes of this paper, we first take the Fourier Transform (FT) of $S_R(x)$ denoted here as $\tilde S_R(k)$. 
It is a well-known result of Fourier analysis that the FT of a convolution is equal to the product of the FTs of each individual function.\footnote{%
See for instance chapter 15 of ref.~\cite{Arfken}. }
With these considerations, the FT $\tilde S_R(k)$ can be written as: 
\begin{align}
\tilde S_R(k) = \sum_{n=0}^{+\infty} P( n; \mu ) \tilde S^n(k)  \tilde B(k). 
\label{eq:ft}
\end{align}
$\tilde S^n(k)$ is the $n$-th power of $\tilde S(k)$, the FT of the SPE response function $S(x)$, and $ \tilde B(k)$ is the FT of the pedestal noise. 
Owing to the mathematical form of the Poisson $P( n; \mu )$ factors, eq.~\eqref{eq:poisson}, the series of eq.~\eqref{eq:ft} can be formally summed and the result is: 
\begin{align}
\tilde S_R(k) =\tilde B(k) e^{\mu ( \tilde S(k) - 1 )}. 
\label{eq:master}
\end{align}
Of course, it is obvious that the Inverse FT of eq.~\eqref{eq:master} gives $S_R(x)$. 

If one could invert the formula of eq.~\eqref{eq:master} a closed form of $S_R(x)$ could then be obtained. 
Unfortunately, this inversion is not always possible in an analytical way. 
In several cases, even a straightforward calculation of $\tilde S(k)$ is not at all trivial. 
To bypass these obstacles, we evaluate FTs (and their inverse) numerically using Discrete Fourier Transform (DFT). 
If the steps chosen in DFT are small enough a good estimation of $S_R(x)$ is achieved. 
Additionally, these operations are fast by virtue of the many excellent Fast Fourier Transform (FFT) algorithms currently available. 

In practice, we employed the well-known C library FFTW~\cite{fftw} for the evaluation of the real ($\Re$) and imaginary ($\Im$) parts of both $\tilde S(k)$ and $\tilde B(k)$.
Eq.~\eqref{eq:master} can then be written as: 
\begin{align}
\tilde S_R(k) = | \tilde B(k) | e^{ \mu ( \Re[ \tilde S(k) ] - 1 ) }   e^{i ( \phi_{\tilde B} + \mu \Im[\tilde S(k) ] )}, 
\label{eq:master2}
\end{align} 
where $\Re[ \tilde S(k) ]$ and $\Im[ \tilde S(k) ]$ are the real and imaginary parts of $\tilde S(k)$, and $| \tilde B(k) |$, $\phi_{\tilde B}$ are the magnitude and argument of $\tilde B(k)$ respectively. 
The complex parts of $\tilde S_R(k)$ are given by:
\begin{align}
\Re[\tilde S_R(k) ] & = | \tilde B(k) |  e^{ \mu ( \Re[ \tilde S(k) ] - 1 ) }  \cos( \phi_{\tilde B }+ \mu \Im[\tilde S(k) ] ) \quad \text{and}, \\
\Im[\tilde S_R(k) ] & =  | \tilde B(k) | e^{ \mu ( \Re[ \tilde S(k) ] - 1 ) }  \sin( \phi_{\tilde B } + \mu \Im[\tilde S(k) ] ). 
\label{eq:parts}
\end{align} 
Inserting $\Re [ \tilde S_R(k) ]$ and  $\Im [ \tilde S_R(k) ]$ in the Inverse DFT (IDFT) implemented in the FFTW library one can then retrieve the values of $S_R(x)$ numerically. 

For all the analyses presented in this article, a step half the size of the histogram binning was used for the DFTs and IDFT. 
This was sufficient to reach the required accuracy level. 
A \texttt{C++/ROOT} based~\cite{root} software that implements the DFT approach and evaluates $S_R(x)$ is available in \cite{git}, a public \texttt{git-hub} repository.
The code is simple, with several examples, and can be used to fit SPE distributions extracting the gain and all the other parameters of $S(x)$. 
It is quite efficient in the sense that it only needs a few seconds to analyze a single distribution.  
Last, we should emphasize that, in contrast to eq.~\eqref{eq:Sr} the DFT method is able to derive $S_R(x)$ for any value of $\mu$ since eq.~\eqref{eq:master} represents the full sum of the $S_R(x)$ series.

\begin{figure}[!t]
\centering
\includegraphics[width=13.2cm, height=10.0cm]{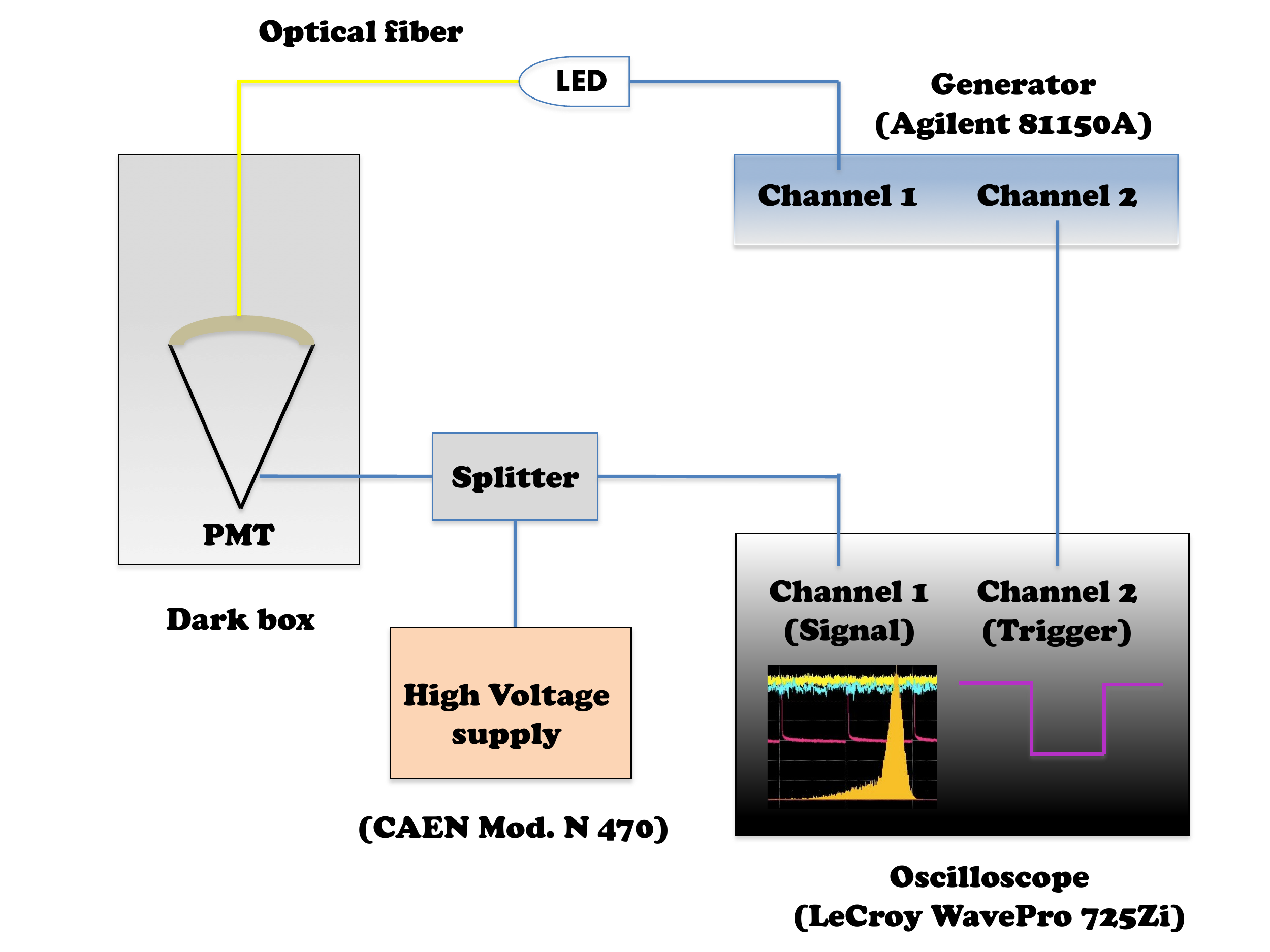} 
\caption{Pictorial representation of our PMT testing apparatus.}
\label{fig:setup}
\end{figure}

\section{Calibration of the Hamamatsu R1408 photomultiplier tube}
\label{sec:results}

In the penultimate section of this article, we showcase the importance and utility of the DFT numerical approach through the absolute gain calibration of the Hamamatsu R1408 PMT model. 
The R1408 PMTs were originally used in the IMB experiment~\cite{IMB} and recently in the Double Chooz experiment~\cite{DC}. 
They are known for not having a sharp peak at the single PE position and their calibration through the original Bellamy method can be complicated.\footnote{%
In IMB the R1408 PMTs were calibrated by means of the occupancy method, ref.~\cite{IMBcalib}. }
Here, we demonstrate that the DFT formalism, coupled with a carefully chosen  $S(x)$ model, can provide rigorous results with an unprecedented precision. 

First steps towards this study were given in ref.~\cite{Me}. 
In this reference the exponential part of $S(x)$ that describes PEs missing the first amplification stage was neglected 
due to the lack of an efficient numerical procedure, and only approximate values for the gains were produced.
This was satisfactory for the needs of the Double Chooz Inner Veto calibration. 
We now show how this small caveat can be patched up; the circle is now complete.

\subsection{Experimental apparatus}

The main features of the experimental apparatus needed for the absolute calibration of PMTs are well-known and, more or less, standard.
Most of the components we utilized for this purpose are graphically depicted in figure~\ref{fig:setup}. 
As it is shown, the PMT was placed inside a dark box. 
Extra black silicon was used to ensure that the box was light-tight and the whole setup operated inside a small air-conditioned dark room. 
The temperature of the room was fixed at 21 $^\circ$C. 

The light pulses used to illuminate the PMT were created by a Light Emitting Diode (LED) connected to a fast pulse generator (Agilent 81150A~\cite{agi}). 
The light was directed inside the box and onto the PMT through an optical quartz fiber (Thorlabs BFH48-600~\cite{fib}). 
A halo-like plastic support, attached to the PMT, was used to ensure that the fiber was always on the center of the photocathode and just touching it. 
All of the measurements were done with blue light, $ \lambda = 475$~nm, and the generator operated at 500~Hz. 
The width of the signal pulses were 20~ns.  
The signal from the PMT was separated by the input high voltage through a splitter and was driven to an oscilloscope LeCroy WavePro 725Zi~\cite{lecroy}. 

\begin{figure}[!t]
\centering
\includegraphics[width=14.0cm, height=8.75cm]{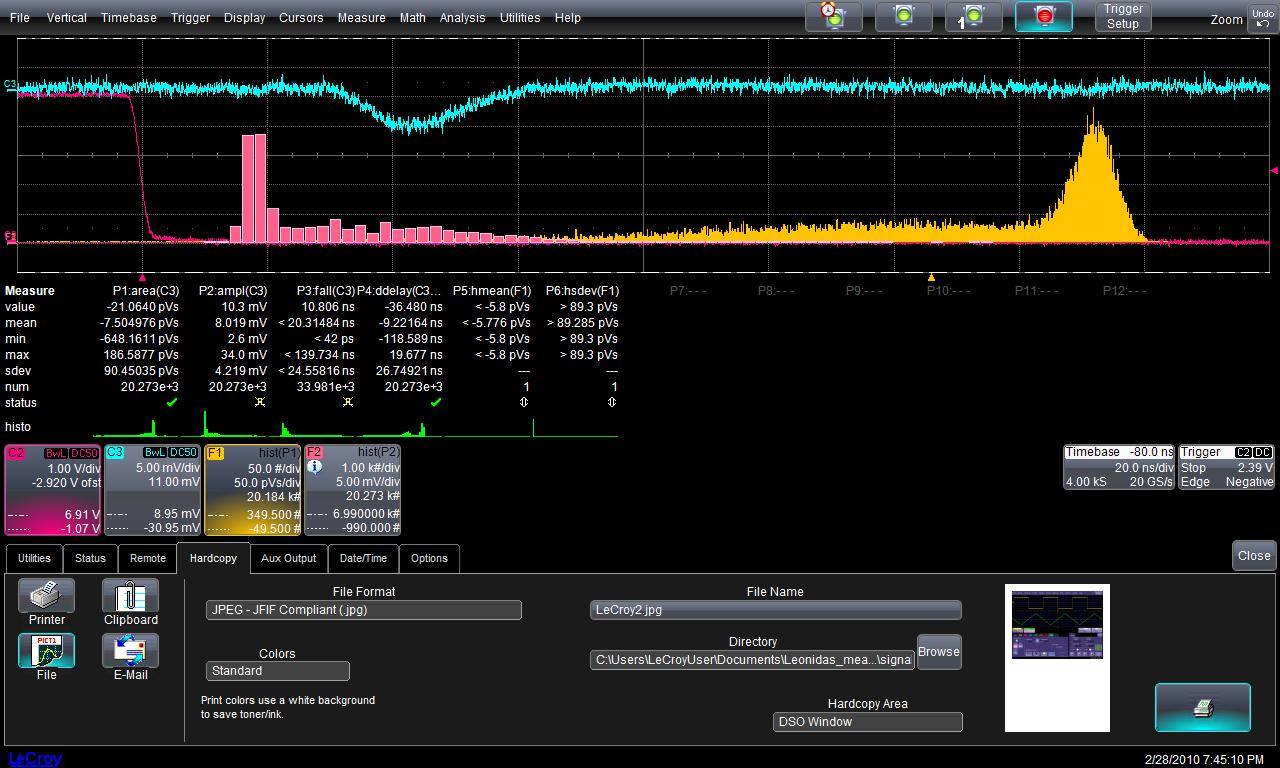} 
\caption{A screenshot from the LeCroy WavePro 725Zi oscilloscope.}
\label{fig:lecroy}
\end{figure}

All data sets were taken with the setup triggering on the generator's second, duplicated, channel sent to the oscilloscope's Channel 2.
This means that no threshold was set in the data acquisition and the charge recorded was integrated in a gate triggered by the generator.  
Figure~\ref{fig:lecroy} shows a screenshot taken from the oscilloscope. 
The purple curve is the trigger sent by the generator and the cyan curve is the signal of the PMT. 
The two histograms, pink and yellow, refer to the voltage and the charge 
measured on the anode of the PMT respectively. 
In what follows we shall concern ourselves exclusively with the output charge on the oscilloscope; that is the yellow histogram. 
Of course, if no additional noise is present the two histograms are analogous as in figure~\ref{fig:lecroy}. 

For the purposes of gain determination the LED input voltage was tuned such that the mean value of the observed PEs would lie within the range of $\mu \simeq 0.1$--$3.0$. 
This setting makes it possible to observe the pedestal peak clearly since then one can easily deconvolute it through the fit. 
Moreover, for higher values of $\mu$ spurious correlations might arise that could bias the results.\footnote{%
See for instance the results of table~1 in ref.~\cite{Bellamy}. }
In any case, it is always more convenient to limit oneselves in this narrow region since the presence of the pedestal gaussian enforces strong constrains to the fit.
In particular it constraints the poissonian mean $\mu$ through the normalization, $e^{-\mu}$, of the pedestal peak.
\enlargethispage{-\baselineskip}

\subsection{Single photoelectron response model}

\begin{figure}[!t]
\centering
\includegraphics[width=10.0cm, height=7.0cm]{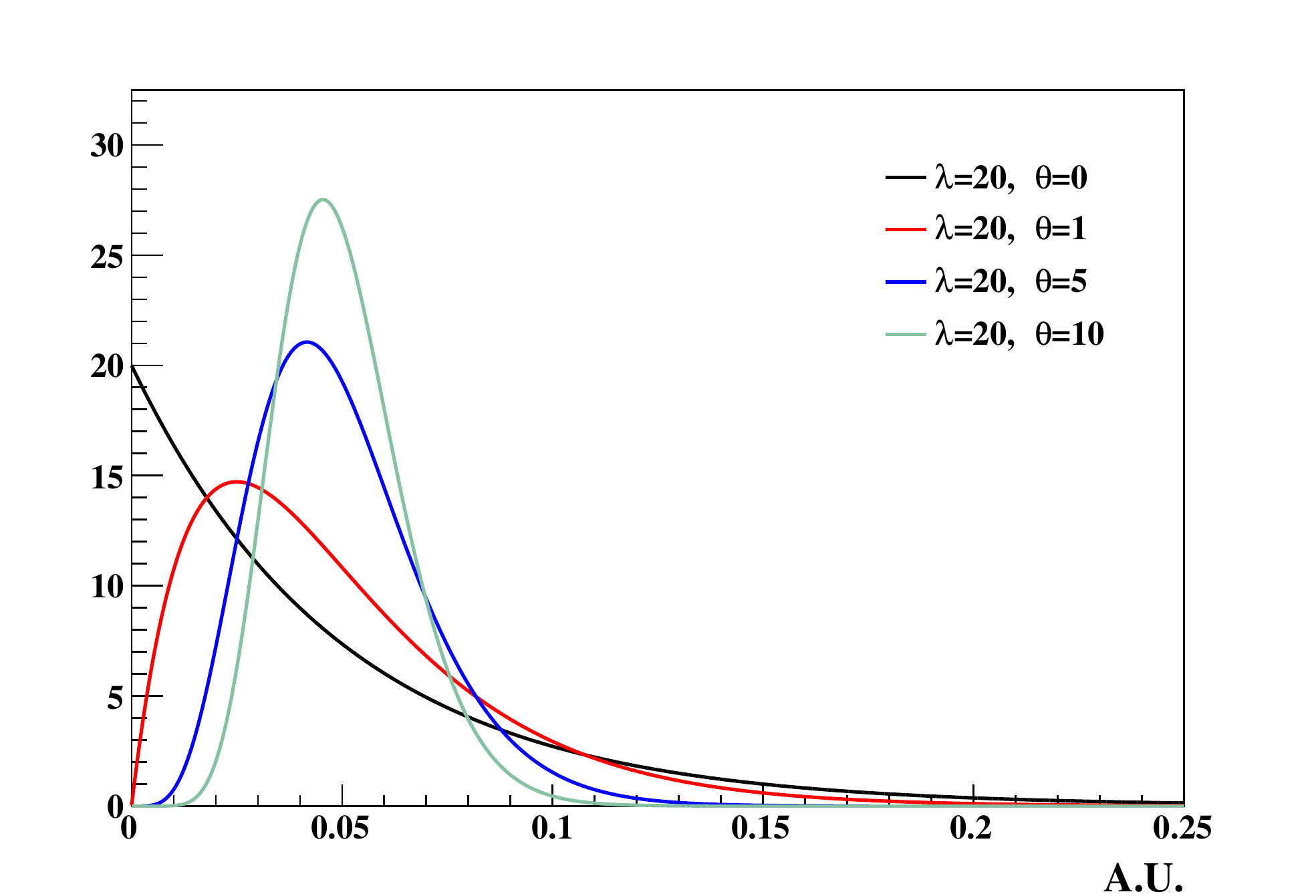} 
\caption{Gamma distribution for $\lambda=20$ and some values of $\theta$. 
As $\theta$ increases the standard deviation of the distribution decreases and its shape becomes more narrow. }
\label{fig:polya}
\end{figure}

The Hamamatsu R1408 8 inches PMTs are difficult to calibrate mainly for two reasons: 
\begin{enumerate}[label=\roman*.]
\item the various PE peaks are entangled due to the significant dispersion of the charge; 
a fact that complicates the extraction of the gain from the fitter\footnote{%
The reader might have a look at figure~3 of ref.~\cite{Bellamy}, where the 2nd PE peak is visible even with naked eye, for an example of a case where there is little charge dispersion.} and, 
\item the SPE response function $S(x)$ cannot be parametrized by a gaussian (truncated or not) and the final model for $S_R(x)$ cannot be solved analytically.    
\end{enumerate}
Fortunately, the gain deconvolution method can still be helpful if one parametrizes the PMT's SPE charge response via the hybrid function:
\begin{align}
S(x) =    \left( \ w \alpha e^{-\alpha x }  +  ( 1 - w ) \lambda (1+ \theta ) \frac{[ \lambda (1+ \theta ) x ]^\theta }{\Gamma (1+\theta) } e^{ - \lambda (1+ \theta ) x } \ \right) H( x ).
\label{eq:polya}
\end{align}
The purely exponential term of eq.~\eqref{eq:polya} (the first term) is absolutely necessary to describe the amplification of PEs that miss the first dynode or photons that convert directly in the first dynode. 
The pre-factor $w$ parametrizes the probability for this to happen. A better discussion of this process is included in ref.~\cite{Smirnov}. 

The second term is a gamma distribution and it describes the full dynode amplification chain process. 
This distribution has been used in the past in connection with the R1408 PMTs calibration~\cite{Me}. 
Note that in this guise, the gamma distribution is also known in the physics vocabulary as \emph{the Polya distribution}. 
Physicists understood its utility in the 60's and 70's in relation to the then newly discovered multiwire gas chambers~\cite{wire}.
$\lambda^{-1}$ is the mean of the gamma distribution and $( \lambda\sqrt{1+\theta} )^{-1}$ its standard deviation. 
When $\theta=0$ the gamma distribution degenerates into a pure exponential. 
On the other hand, when $\theta$ increases the standard deviation decreases and the distribution acquires a more narrow and peaked shape. 
Figure~\ref{fig:polya} shows a few curves for some random values of $\lambda$ and $\theta$. 
In this respect, this model is rather general and it can be used to describe a vast range of PMTs. 

The final model of $S_R(x)$ was solved numerically using the DFT machinery developed in previous section
since the mathematics involved in the calculations of the $S_n(x)$ and $(S_n*B)(x)$ convolutions cannot be solved analytically. 
$S_R(x)$ was then used to fit charge spectra taken with a R1408 PMT inside the dark box of the apparatus and illuminated with low-intensity light pulses. 
A standard gaussian $\chi^2$ function that compares data to the model was built and the global minimum (the so-called best fit) was found using the Minuit2 package~\cite{minuit2}. 
The best fit returns the Poisson mean of the light source ($\mu$), the mean and standard deviation of the pedestal ($Q_0$ and $\sigma_0$) and the four parameters of the $S(x)$ distribution ($w$, $\alpha$, $\lambda$ and $\theta$). 
Note that for the studies included in this article, the charge was measured in nanovolt times second (nVs) which is the unit given by the LeCroy oscilloscope. 

\begin{figure}[!t]
\centering
\includegraphics[width=9.2cm, height=6.5cm]{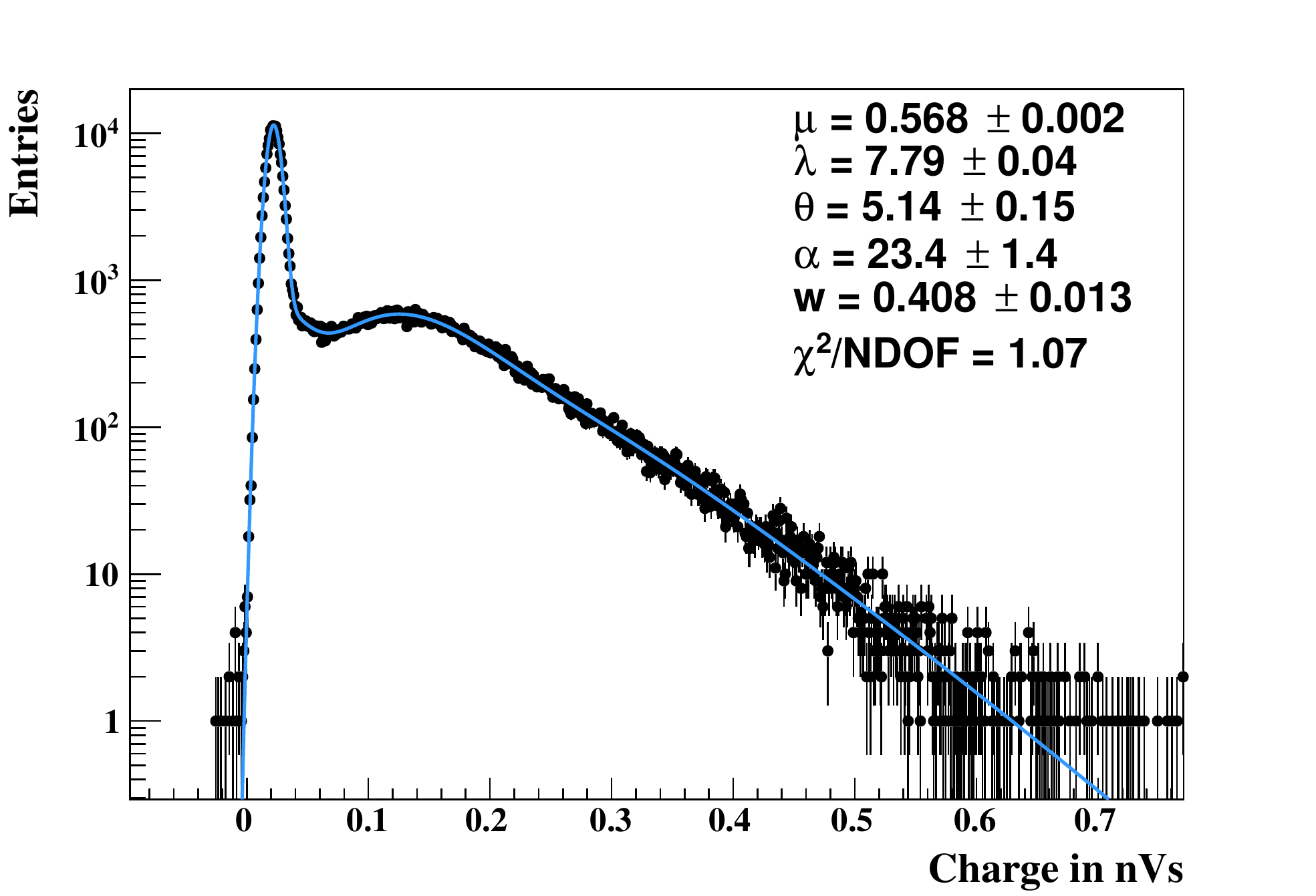} \\[1.5ex]
\includegraphics[width=9.2cm, height=6.5cm]{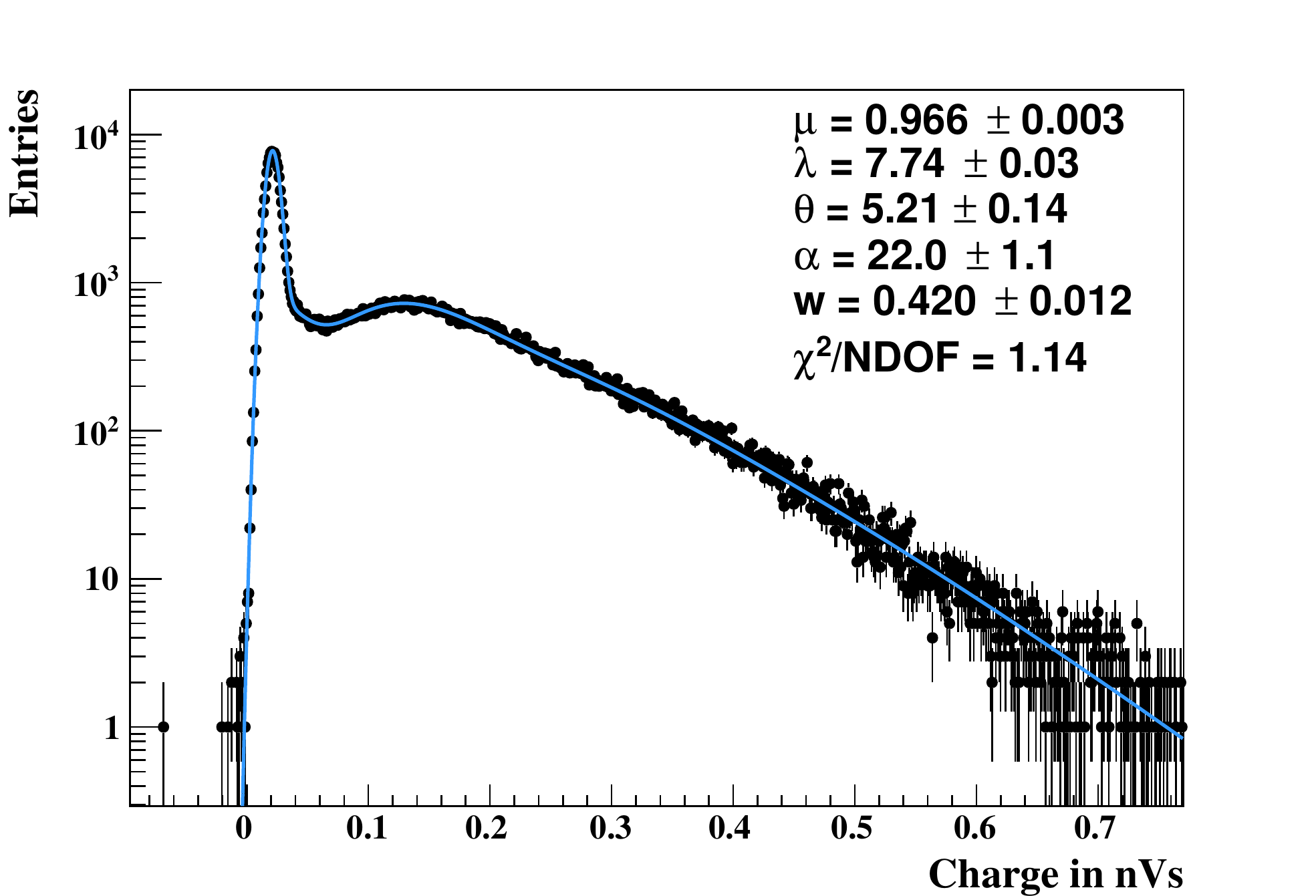}  \\[1.5ex] 
\includegraphics[width=9.2cm, height=6.5cm]{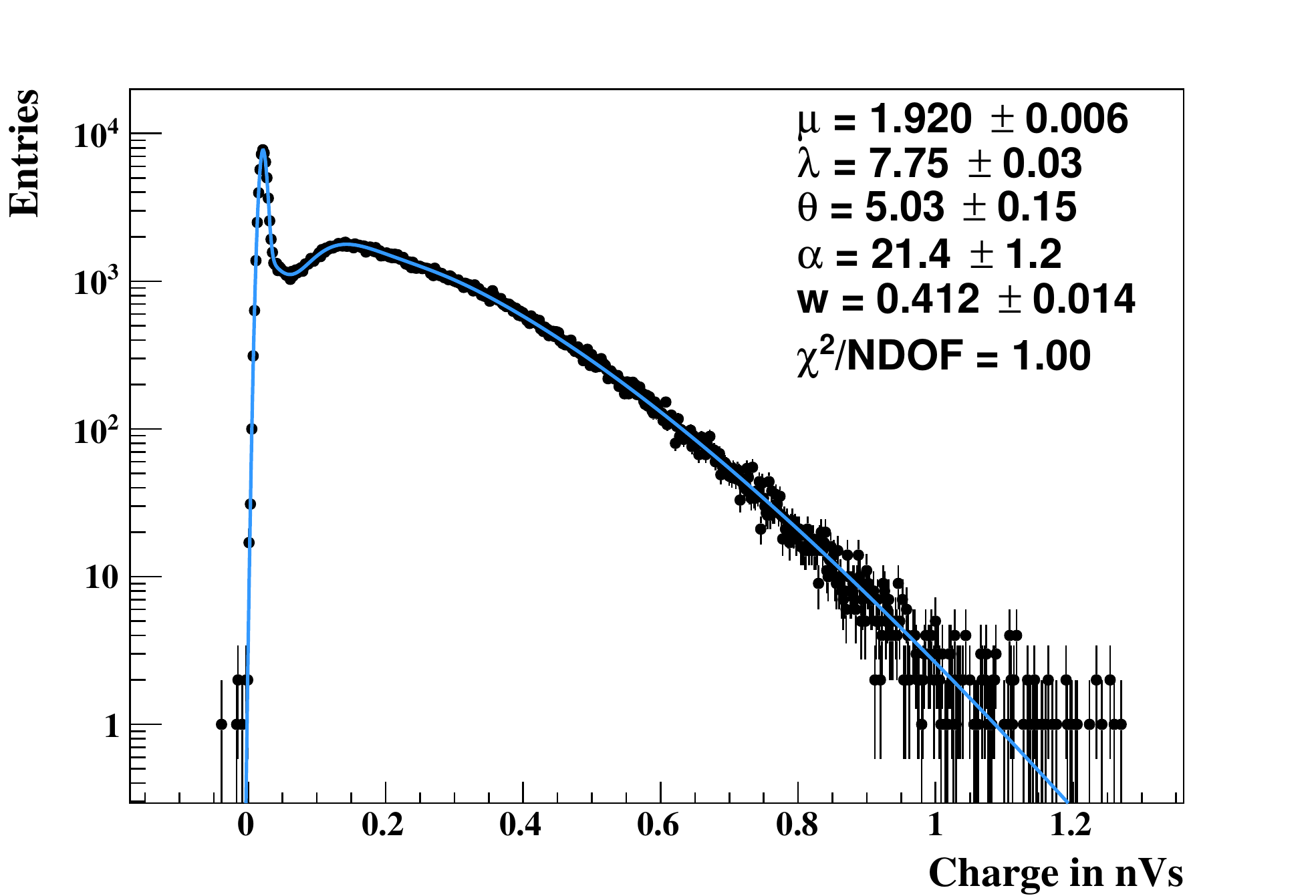} 
\caption{A few fits obtained using the $S(x)$ of eq.~\eqref{eq:polya}. 
The data are shown in the black dots and the best fit curve is shown in azure line. }
\label{fig:fit}
\end{figure}

Figure~\ref{fig:fit} shows  a few fits obtained with the $S_R(x)$ model put forward in this section. 
The best fit curve (azure line) follows closely the data and the $\chi^2$ over the number of degrees of freedom (NDOF) is always close to 1.0. 
For all the histograms treated in this study, a simple gaussian fit was first performed around the maximum of the distribution to extract the mean ($Q_0$) and standard deviation ($\sigma_0$) of the pedestal. 
$Q_0$ and $\sigma_0$ were then fixed within a $\pm$ 2.5\% tolerance. 
Additionally, the fit around the pedestal provided the normalization of the zero PE peak ($N_0$). 
An estimate of the mean number of PEs was given by the formula: 
\begin{align}
\mu \simeq - \ln \left(  \frac{N_0}{\ N_{tot}} \right),  
\label{eq:muestim}
\end{align}
and was input as an initial value to aid the minimizer. 
$N_{tot}$ is the total number of entries and eq.~\eqref{eq:muestim} is obtained from $P(0; \mu ) = e^{-\mu}$. 

The initial values of $\alpha$ and $\lambda$ were set to: 
\begin{align}
\frac{\mu}{ \bar x - Q_0 },
\label{eq:laestim}
\end{align}
where $\bar x$ is the mean of the charge histogram. 
To understand this choice, we note that the average value of $S(x)$ (that is the gain, $G$) is equal to: 
\begin{align}
G = \frac{w}{\alpha} + \frac{1-w}{\lambda}. 
\label{eq:G}
\end{align}
$\alpha^{-1}$ is the mean of the exponential part and $\lambda^{-1}$ the mean of the gamma distribution. 
On the other hand, the average of $S_R(x)$ can be calculated using eq.~\eqref{eq:Sr0} to be:
\begin{align}
\bar S_R = Q_0 + \mu G.  
\label{eq:x}
\end{align}
Approximating $\bar x$ with $\bar S_R$ one can the derive the formula: 
\begin{align}
G \simeq \frac{\bar x - Q_0}{\mu}. 
\end{align}
We see that putting $\alpha$ and $\lambda$ equal to $G^{-1}$ one gets the correct order of magnitude.  

Finally, $w$ was first set to 0.2 and limited between 0 and 0.6 (since for every well-functioning PMT the probability of badly amplified PEs cannot be larger than 60\%). 
On the other hand, $\theta$ was initialized to 7 and constrained between the large limits 0.7 -- 56. 
The standard deviation of the gamma distribution is: 
\begin{align}
\sigma = \frac{1}{\lambda\sqrt{1+\theta}}. 
\end{align}
Letting $\theta = 7$, it yields a relative $\sigma$ over the $\lambda^{-1}$ mean of $\sim$ 35\% which is close to what most PMTs have. 
It should be emphasized that the aforementioned limits and initial values are absolutely necessary. 
The multidimensional fit of $S_R(x)$ is very complicated and it is likely to fail unless good initial values are inserted in the minimizer. 
Note also that the fit, when successful, subtracts \emph{statistically} the pedestal through the determination of $Q_0$, $\sigma_0$ and allows for a precise evaluation of all the parameters of $S(x)$. 

\subsection{Results}

\begin{table}[t!]
\centering
\begin{tabular}{| c  || c | c | c | c || c |}
\hline
$\mu$  &  $w$ &  $\alpha$ &  $\lambda$ &  $\theta$ & $\chi^2$/NDOF\\[0.6ex] \hline\hline
0.568 $\pm$ 0.002 & 0.408 $\pm$ 0.013 & 23.4 $\pm$ 1.4 & 7.79 $\pm$ 0.04 & 5.14 $\pm$ 0.15 & 1.07 \\
0.626 $\pm$ 0.002 & 0.430 $\pm$ 0.012 & 21.0 $\pm$ 1.0 & 7.78 $\pm$ 0.03 & 5.15 $\pm$ 0.14 & 1.28 \\
0.692 $\pm$ 0.003 & 0.434 $\pm$ 0.013 & 20.9 $\pm$ 1.1 & 7.70 $\pm$ 0.03 & 5.41 $\pm$ 0.16 & 0.95 \\
0.722 $\pm$ 0.002 & 0.413 $\pm$ 0.011 & 22.5 $\pm$ 1.1 & 7.77 $\pm$ 0.03 & 5.10 $\pm$ 0.12 & 1.21 \\
0.819 $\pm$ 0.003 & 0.405 $\pm$ 0.012 & 23.4 $\pm$ 1.2 & 7.79 $\pm$ 0.03 & 5.03 $\pm$ 0.13 & 1.18 \\
0.924 $\pm$ 0.003 & 0.415 $\pm$ 0.011 & 22.1 $\pm$ 1.0 & 7.75 $\pm$ 0.02 & 5.10 $\pm$ 0.12 & 1.08 \\
0.966 $\pm$ 0.003 & 0.420 $\pm$ 0.012 & 22.0 $\pm$ 1.1 & 7.74 $\pm$ 0.03 & 5.21 $\pm$ 0.14 & 1.14 \\
1.121 $\pm$ 0.004 & 0.422 $\pm$ 0.011 & 21.5 $\pm$ 1.0 & 7.73 $\pm$ 0.03 & 5.23 $\pm$ 0.13 & 1.08 \\
1.309 $\pm$ 0.004 & 0.430 $\pm$ 0.014 & 20.7 $\pm$ 1.1 & 7.75 $\pm$ 0.03 & 5.21 $\pm$ 0.16 & 1.08 \\
1.379 $\pm$ 0.004 & 0.417 $\pm$ 0.012 & 22.4 $\pm$ 1.1 & 7.74 $\pm$ 0.03 & 5.20 $\pm$ 0.14 & 1.17 \\
1.645 $\pm$ 0.005 & 0.404 $\pm$ 0.012 & 23.2 $\pm$ 1.3 & 7.77 $\pm$ 0.03 & 5.00 $\pm$ 0.13 & 1.16 \\
1.920 $\pm$ 0.006 & 0.412 $\pm$ 0.014 & 21.4 $\pm$ 1.2 & 7.75 $\pm$ 0.03 & 5.03 $\pm$ 0.15 & 1.00 \\
2.113 $\pm$ 0.006 & 0.402 $\pm$ 0.013 & 21.9 $\pm$ 1.2 & 7.75 $\pm$ 0.03 & 4.99 $\pm$ 0.14 & 1.13 \\
2.549 $\pm$ 0.010 & 0.424 $\pm$ 0.017 & 20.1 $\pm$ 1.2 & 7.74 $\pm$ 0.03 & 5.34 $\pm$ 0.22 & 1.24
\\[0.6ex] \hline\hline
\end{tabular}
\caption{Summary of the R1408 PMT calibration results.}
\label{tab:money}
\end{table}

To demonstrate that the fitting model is self-consistent and that extracts the correct $w$, $\alpha$, $\lambda$ and $\theta$, several data were taken with varying LED intensity. 
The results of the analysis are shown in table~\ref{tab:money}. 
As the same PMT is used, the fit should output compatible values for all the parameters of $S(x)$ regardless of $\mu$. 
The pedestal might drift or change, depending on the stability of the setup, but the SPE response should stay the same. 

Indeed, in table~\ref{tab:money} one can see that as $\mu$ becomes larger $w$, $\alpha$, $\lambda$ and $\theta$ remain within the errors. 
A few further comments should be made.  
First, $\lambda$ can be obtained with much better accuracy than $w$, $\alpha$ and $\theta$. 
This is the case because the contribution of misamplified PEs is mainly constrained from the valley between the pedestal and the SPE peak. 
It appears that the minimizer has more difficulty to identify $w$ and $\alpha$ in this narrow region. 
Nonetheless, consistent results can be obtained. 
In particular, the pre-factor $w$ (the probability that PEs are not amplified in the full dynode chain) stays constant. 
This should be contrasted with ref.~\cite{Bellamy} where the exponential part of $S(x)$ is attached to the pedestal and $w$ increases with light intensity. 

\begin{figure}[!t]
\centering
\includegraphics[width=10.0cm, height=6.8cm]{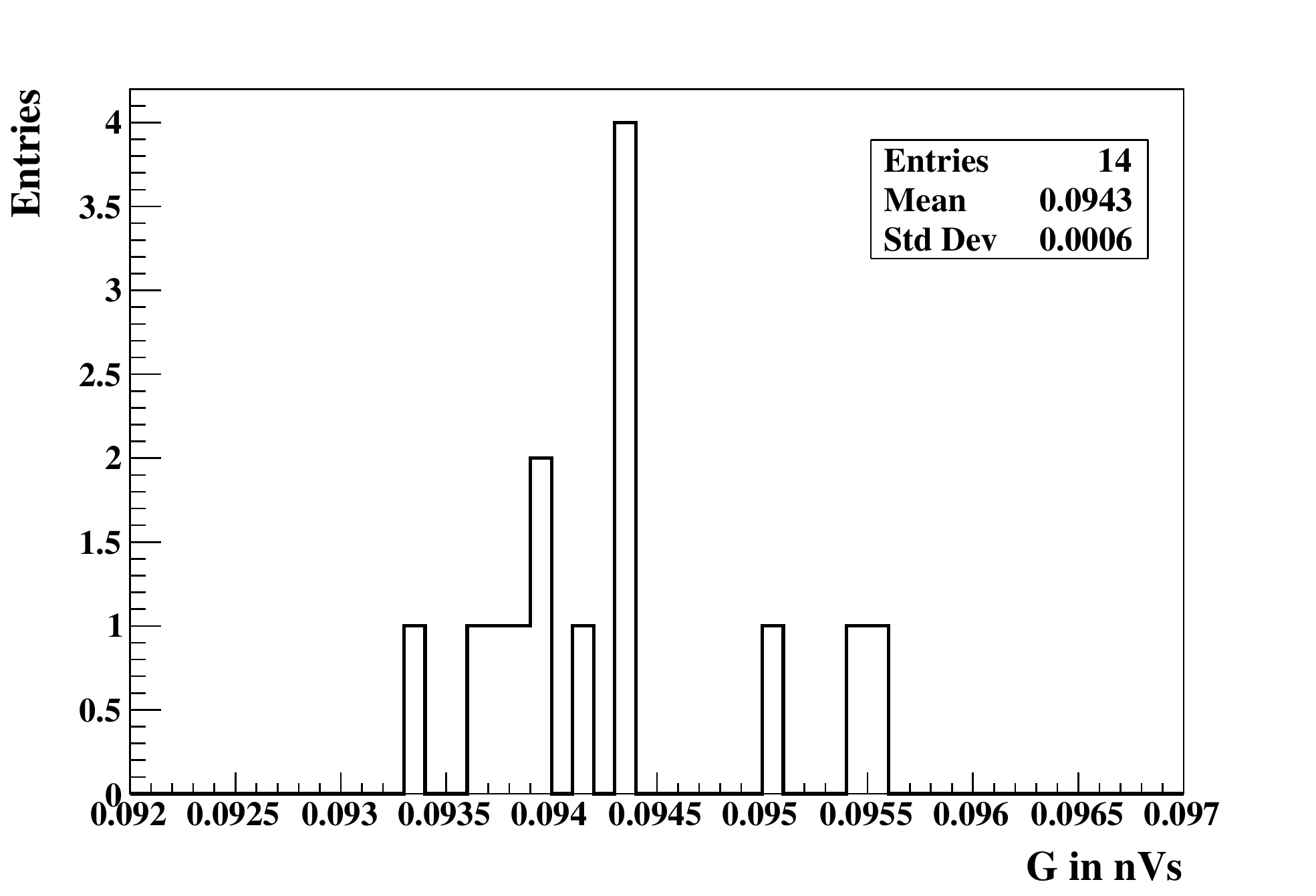} 
\caption{Distribution of the gain ($G$) for the measurements compiled in table~\ref{tab:money}. }
\label{fig:g}
\end{figure}

Second, even though correlations exist between the various parameters of $S(x)$, 
a straightforward calculation of the gain ($G$) using eq.~\eqref{eq:G}, gives results that are consistent. 
Figure~\ref{fig:g} shows the one-dimensional distribution of $G$ for those measurements included in table~\ref{tab:money}. 
The relative standard deviation of the distribution in figure~\ref{fig:g} is very small, highlighting that one can determine $G$ with a $\sim$1\% accuracy. 
These are important results, indicating that the gamma function SPE model can be useful in the absolute calibration of PMTs providing precise results in a vast range of PEs.


\section{Conclusions}
\label{sec:last}

\newcommand{\PMT}{PMT}
\newcommand{\SPE}{SPE}
\newcommand{\pdf}{probability density function}

In this paper we have presented a novel method for calibration of photomultiplier tubes.
This method is particularly effective for modeling \PMT{}s that do not present regular characteristics, where traditional methods yield acceptable results.
With our proposed method any \SPE{} \pdf{} can be used to characterize the \PMT{} response, as long as such function can be efficiently calculated numerically.
This is possible thanks to the usage of efficient Fourier Transformation algorithms available, which are then used to calculate the many required convolutions of the \SPE{} \pdf{}
with itself and with the pedestal noise in order to represent the multi-photoelectron \pdf{}s.

The proposed method was successfully used on a R1408 Hamamatsu PMT, where the \SPE{} \pdf{} was taken by adding an exponential and a gamma (or Polya) distribution, both multiplied by a Heavyside function.
This \SPE{} \pdf{} description used to characterize this \PMT{} is too complex to be used with brute force numerical convolutions, and it would typically need to be simplified before being used.
With our method, this function can be used efficiently without any additional simplification for fitting the \PMT{}'s \SPE{} \pdf{} parameters using a usual LED calibration setup.
It was verified that the \PMT{} parameters obtained are stable for varying incident light intensity, and in all cases a $\chi^2/\text{NDOF}$ close to one is obtained for the fits.
A gain determination accuracy of $\sim1\%$ was achieved with this procedure for this \PMT{}, for a light source mean $\mu$ between $\sim$ 0.1--3.0. 


%

%


\begin{thebibliography}{99}
%

\bibitem{Leo} W. R. Leo, \emph{Techniques for Nuclear and Particle Physics Experiments: A How-to Approach}, 2nd ed. Springer, Berlin, Germany (1994).

\bibitem{SK} S. Fukuda {et al.} (Super-Kamiokande Collaboration), \emph{The Super-Kamiokande detector}, 
\href{https://www.sciencedirect.com/science/article/pii/S016890020300425X}{\emph{Nucl. Instrum. Meth.} {\bf A 501} (2003) 418-462}.

\bibitem{JUNO1}  Z. Djurcic {et al.} (JUNO Collaboration), \emph{JUNO Conceptual Design Report}, (2015) [\href{https://arxiv.org/abs/1508.07166}{\texttt{arXiv:1508.07166}}].

\bibitem{JUNO2} F. An {et al.} (JUNO Collaboration), \emph{Neutrino Physics with JUNO}, (2015) [\href{https://arxiv.org/abs/1507.05613}{\texttt{arXiv:1507.05613}}]. 

\bibitem{JUNO3} M. He (for the JUNO Collaboration), \emph{Double Calorimetry System in JUNO}, \emph{in TIPP 2017 proceeding} (2017) [\href{https://arxiv.org/abs/1706.08761}{\texttt{arXiv:1706.08761}}]. 




\bibitem{Bellamy} E. H. Bellamy {et al.}, \emph{Absolute Calibration and Monitoring of a spectrometric channel using a photomultiplier}, 
\href{https://www.sciencedirect.com/science/article/pii/016890029490183X}{\emph{Nucl. Instrum. Meth.} {\bf A 339} (1994) 468-476}. 




\bibitem{DCID1} E. Calvo {et al.}, \emph{Characterization of large area photomutipliers under low magnetic fields: design and performances of the magnetic shielding for the Double Chooz neutrino experiment}, 
\href{https://www.sciencedirect.com/science/article/pii/S016890021001199X}{\emph{Nucl. Instrum. Meth.} {\bf A 621} (2010) 222-230} 
 [\href{https://arxiv.org/abs/0905.3246}{\texttt{arXiv:0905.3246}}]. 

\bibitem{DCID2} C. Bauer {et al.}, \emph{Qualification Tests of 474 Photomultiplier Tubes for the Inner Detector of the Double Chooz Experiment}, 
\href{https://iopscience.iop.org/article/10.1088/1748-0221/6/06/P06008}{\emph{JINST} {\bf 6} (2011) P06008}  [\href{https://arxiv.org/abs/1104.0758}{\texttt{arXiv:1104.0758}}]. 

\bibitem{DCID3} T. Matsubara {et al.}, \emph{Evaluation of 400 low background 10-in. photo-multiplier tubes for the Double Chooz experiment}, 
\href{https://www.sciencedirect.com/science/article/pii/S016890021101775X?via\%3Dihub}{\emph{Nucl. Instrum. Meth.} {\bf A 661} (2012) 16-25} 
[\href{https://arxiv.org/abs/1104.0786}{\texttt{arXiv:1104.0786}}]. 


\bibitem{Me} L. N. Kalousis, \emph{Calibration of the Double Chooz detector and cosmic background studies}, \href{http://inspirehep.net/record/1295030}{\emph{PhD thesis, University of Strasbourg} (2012)}. 
%



\bibitem{Hama} Hamamatsu, \emph{Photomultiplier tubes and assemblies for scintillation counting \& High Energy Physics} 

\href{https://www.hamamatsu.com/resources/pdf/etd/High_energy_PMT_TPMTZ0003E.pdf}{https://www.hamamatsu.com/resources/pdf/etd/High\_energy\_PMT\_TPMTZ0003E.pdf}

\bibitem{Zorin} I. Chirikov-Zorin {et al.}, \emph{Method for precise analysis of the metal package photomultiplier single photoelectron spectra}, 
\href{https://www.sciencedirect.com/science/article/pii/S0168900200005933}{\emph{Nucl. Instrum. Meth.} {\bf A 456} (2001) 310-324}.


\bibitem{Smirnov}  R. Dossi {et al.}, \emph{Methods for precise photoelectron counting with photomultipliers}, 
\href{https://www.sciencedirect.com/science/article/pii/S0168900200003375}{\emph{Nucl. Instrum. Meth.} {\bf A 451} (2000) 623-637}. 

\bibitem{darkside} T. Alexander {et al.} (DarkSide Collaboration), \emph{Light Yield in DarkSide-10: a Prototype Two-phase Argon TPC for Dark Matter Searches}, 
\href{https://www.sciencedirect.com/science/article/pii/S0927650513001254?via\%3Dihub}{\emph{Astropart. Phys.} {\bf 49} (2013) 44-51}  
[\href{https://arxiv.org/abs/1204.6218}{\texttt{arXiv:1204.6218}}]. 

\bibitem{pavel} P. Degtiarenko, \emph{Precision analysis of the photomultiplier response to ultra low signals}, 
\href{https://www.sciencedirect.com/science/article/pii/S0168900217308252?via\%3Dihub}{\emph{Nucl. Instrum. Meth.} {\bf A 872} (2017) 1-15} 
[\href{https://arxiv.org/abs/1608.07525}{\texttt{arXiv:1608.07525}}]. 



\bibitem{Arfken} G. B. Arfken and H. J. Weber, \emph{Mathematical Methods for Physicists}, 6th Ed. Elsevier Academic Press, San Diego, USA (2005).

\bibitem{fftw} \href{http://www.fftw.org}{http://www.fftw.org} 

\bibitem{root} R. Brun and F. Rademakers, \emph{ROOT - An Object Oriented Data Analysis Framework}, \emph{Proceedings AIHENP'96 Workshop}, Lausanne Switzerland (1996),
\href{https://www.sciencedirect.com/science/article/pii/S016890029700048X?via\%3Dihub}{\emph{Nucl. Instrum. Meth.} {\bf A 389 } (1997) 81-86.} \\
See also \href{http://root.cern.ch/}{http://root.cern.ch/}

\bibitem{git} \href{https://github.com/kalousis/PMTCalib}{https://github.com/kalousis/PMPMTCalib}

\bibitem{IMB} R. Becker-Szendy {et al.} (IMB Collaboration), \emph{IMB-3 : a large water Cherenkov detector for nucleon decay and neutrino interactions}, 
\href{https://www.sciencedirect.com/science/article/pii/016890029390998W?via\%3Dihub}{\emph{Nucl. Instrum. Meth.} {\bf A 324} (1993) 363-382}.

\bibitem{DC} Y. Abe  {et al.} (Double Chooz Collaboration), \emph{Reactor $\bar\nu_e$ disappearance in the Double Chooz experiment}, 
\href{https://doi.org/10.1103/PhysRevLett.108.131801}{\emph{Phys. Rev.} {\bf D}  86 (2012) 052008}    [\href{https://arxiv.org/abs/1207.6632}{\texttt{arXiv:1207.6632}}]. 

\bibitem{IMBcalib} R. Becker-Szendy {et al.} (IMB Collaboration), \emph{Calibration of the IMB detector}, 
\href{https://www.sciencedirect.com/science/article/pii/0168900295900187?via\%3Dihub}{\emph{Nucl. Instrum. Meth.} {\bf A~352} (1995) 629-639}.

\bibitem{agi} \href{https://www.keysight.com/en/pd-1287544-pn-81150A/pulse-function-arbitrary-noise-generator?nid=-536902255.748669.00&cc=FR&lc=fre}{https://www.keysight.com/en/pd-1287544-pn-81150A}

\bibitem{fib} \href{https://www.thorlabs.com/thorproduct.cfm?partnumber=BFH48-600}{https://www.thorlabs.com/thorproduct.cfm?partnumber=BFH48-600}

\bibitem{lecroy} \href{http://cdn.teledynelecroy.com/files/pdf/wavepro_7_zi-a_datasheet.pdf}{http://cdn.teledynelecroy.com/files/pdf/wavepro\_7\_zi-a\_datasheet.pdf}

\bibitem{wire} V. Fran\c{c}ais, \emph{Description and simulation of the physics of Resistive Plate Chambers},  
 \href{https://tel.archives-ouvertes.fr/tel-01727712}{\emph{PhD thesis, University of Clermont Auvergne} (2017)}. 

\bibitem{minuit2} F. James and M. Winkler, \emph{C++ MINUIT User's Guide},
\href{https://root.cern.ch/root/htmldoc/guides/minuit2/Minuit2.html}{https://root.cern.ch/root/htmldoc/guides/minuit2/Minuit2.html}


%


\end{thebibliography}
\end{document}